\documentclass[twocolumn]{revtex4-2} 
\usepackage{hyperref}
\usepackage{tabularx}
\usepackage{multirow}
\usepackage{graphicx}
\usepackage{dcolumn}
\usepackage{amsmath} 
\usepackage{mathrsfs} 
 
\begin{document}

\preprint{APS/123-QED}

\title{Constructing Maximal Extensions of the Vaidya Metric in Israel Coordinates: \texorpdfstring{\\}{\textbackslash}
 I. Integration of the Field Equations}

\author{Sheref Nasereldin}
 \email{16sna1@queensu.ca}
\author{Kayll Lake}%
 \email{lakek@queensu.ca}
\affiliation{%
 Department of Physics, Queen's University, Kingston, Ontario, Canada, K7L3N6
}
\date{\today}
\begin{abstract}
This paper explores a complete representation of the Vaidya model, a radial flux of radiation in the eikonal approximation, used for modeling various phenomena in both classical and semi-classical General Relativity and Astrophysics. The majority of the applications of the Vaidya model have been formulated in an incomplete representation. A complete representation is obtained here by direct integration of the Einstein field equations. We  present the methodology to obtain this complete representation, and its utility in the modeling of general relativistic phenomena.
\end{abstract}
\maketitle
\section{Introduction}\label{Sec:Introduction}
The Schwarzschild metric \citep{Schwarzschild1916} has been used to study the exterior geometry of spherical stellar objects undergoing gravitational collapse \cite{PhysRev.56.455,Dewitt1964RelativityGA}, where it is assumed that the radiation emitted by the object is insignificant. However, during the advanced stages of stellar collapse, these objects are expected to emit a considerable amount of mass in the form of radiation, see for example \cite{1965_Harrsion_Thorne_Wheeler}. Therefore, the exterior of a collapsing stellar object is no longer empty, and the Schwarzschild vacuum metric is no longer suitable for its description. The Vaidya metric \cite{Vaidya_II,Vaidya_III} is more suitable for this situation and has been widely used to classically study the geometry outside \footnote{With suitable boundary conditions, such as Israel's conditions, see \cite{1966NCimB..44....1I}, on the spherical surface, this exterior solution can be matched to some proper interior solution, see for example \cite{Vaidya1966AnAS} and \cite{Fayos_2008}.} radiating spherical stellar objects, see for example \cite{PhysRevD.31.233,Castagnino_Umerez,PhysRevD.25.2527,PhysRevD.22.2305,PhysRevD.19.2838, Hamity_Gleiser,PhysRevD.24.3019,adams1994analytic,1988MNRAS.231.1011D,PhysRevD.70.084004,PhysRevD.74.044001}. 
 Thus, one can treat this dynamical mass distribution with its envelope of radiation as an isolated system existing in otherwise vacuum, asymptotically flat spacetime that is described by the Schwarzschild vacuum metric. 

The ``self-similar" Vaidya metric has been used to construct spacetimes that exhibit a visible strong singularity, demonstrating the potential for the failure of the Penrose ``Cosmic censorship hypothesis" \cite{Cosmic_censorship}. This conjecture states that singularities arising from regular initial conditions do not have any causal influence on spacetime. If the hypothesis were to fail, it would be a major flaw in the theory of general relativity and would make it impossible to predict the events in any region of spacetime containing a singularity, as new information could emerge in an unpredictable manner. The growth of curvature along non-spacelike geodesics has been examined (see for example, \cite{Hiscock_Williams_Eardley,papapetrou1985random,1986CQGra...3L.111H,Wurster,PhysRevLett.59.2137,Demianski_Lasota,King_Lasota,PhysRevD.35.1531,PhysRevD.24.3019,PhysRevD.26.518,PhysRevD.26.1479,Santos_singularity,dwivedi1989nature,PhysRevLett.68.3129,dwivedi1994occurrence,Joshi,Hollier}), and the visible singularity in self-similar spacetimes has been classified as strong. Furthermore, Lake and Zannias \cite{PhysRevD.41.3866} showed that the emergence of naked singularities in these spacetimes is due to the self-similarity assumption, rather than spherical symmetry. 

On the semi-classical level, the Vaidya metric has been utilized to explore black hole evaporation, possibly due to Hawking's radiation \cite{Hawking_BH_evaporating}, (see for example \cite{1982NCimB..70..201B,Kuroda,zhen_Zhao,beciu1984evaporating,PhysRevD.23.2823,Sung_Won,Hiscock_William,PhysRevD.41.1356,PhysRevD.63.041503,2006GReGr..38..425F,KAWAI,PhysRevD.91.044020,Balbinot1990}). Furthermore, the Vaidya metric in the double-null coordinates (the mass function must be linear) \cite{Vaidya_DoubleNull} has been used to study the quasi-normal modes (QNM) as a model that supposedly will give deeper insights on the gravitational excitations of black holes (see for example \cite{abdalla2006quasinormal}).

Despite the fact that the majority of applications were structured with the Vaidya metric written in the Eddington-Finkelstein-Like (EFL) coordinates, these coordinates have been known for some time to be incomplete (see for example \cite{Lindquist1965,Israel1967}), leaving the Vaidya manifold not maximally covered. Thus, to ensure the accuracy of all applications, it is required to construct a complete set of coordinates and thoroughly assess the impact of this set of coordinates. This is the primary objective of this paper. In a separate manuscript \cite{Null_Israel_Completeness}, we introduce explicit mass functions as candidates for three distinct Vaidya models. Moreover, we assess the completeness of Israel coordinates in relation to these mass functions.

We organize this paper as follows. In the next section, we review the EFL coordinates and provide a proof of  incompleteness of this set of coordinates, which is the main motivation for any subsequent coordinate representation. In Section \ref{Sec:Israel}, we review the use of Israel coordinates \cite{Israel1966} to write the Vaidya metric \cite{Israel1967}, and discuss why the derivation of these coordinates resulted in unsatisfactory results when attempting to obtain maximal coverings of the Vaidya manifold. The main results of this paper are outlined in Section \ref{Sec:Construction}, in which we introduce an algorithmic method to obtain Israel coordinates by direct integration of the field equations, without relying on any coordinate transformation. In Section \ref{Sec:Restrictions}, we present necessary physical restrictions that must be imposed on the flux of radiation. In Section \ref{Sec:Horizons}, we provide a general derivation regarding the location of the apparent horizon in the Vaidya manifold. It is emphasized that the location of the apparent horizon is established before introducing any expressions to the characterizing functions. In Section \ref{Sec:Representations}, we demonstrate that our construction can be used to obtain both EFL and Israel coordinates by choosing different expressions for the functions that arise from integrating the field equations; such functions, as well as the coefficient of the cross term in the general metric that is presented, shall be referred to as the ``characterizing functions". In Section \ref{Sec:Invariants}, we briefly calculate some of the invariants of the Vaidya metric in Israel coordinates. The last section highlights the main results of the paper and discusses the possible extensions of the current work.
\section{The EFL Coordinates} \label{Sec:EFL}
The Vaidya metric, in the EFL coordinates, is a spherically symmetric solution to the Einstein field equations with the energy momentum tensor approximated in ``the eikonal form" \cite{MTW,Landau_Lifshitz}, which expresses a unidirectional radial flow of unpolarized radiation, 
\begin{equation}
    T_{\alpha\beta} = \Phi k_{\alpha}k_{\beta}= \frac{\epsilon}{4\pi r^2}\frac{dm(u)}{du}k_{\alpha}k_{\beta},
\end{equation}
where $\epsilon = \pm 1 $ and $k_\alpha = \delta^u_\alpha$ is tangent to radial inward or outward-going null geodesics. The spacetime line element in the EFL coordinates takes the form
\begin{equation}\label{Vaidya_EF_coords}
    ds^2 = -\left(1-\frac{2m(u)}{r}\right)du^2+2\epsilon dudr+r^2d\Omega^{2}_{2},
\end{equation}
where $d\Omega^{2}_{2} = d\theta^{2}+sin^2\theta d\phi^2$ is the metric of a unit 2-sphere. For $\epsilon = +1$, the metric expresses inward-directed radiation (towards smaller values of the radius $r$) with a monotonically increasing $m$ as a function of the ``advanced time" coordinate $u$. If $\epsilon = -1$, the metric is that of outgoing radiation (towards larger values of the radius $r$) with $m$ being monotonically decreasing as a function of the ``retarded time" coordinate $u$. However, it is conventional, as stated in \cite{Poisson2004,Griffiths:2009dfa,hawking1973large}, to assign $u$ as the retarded time and $v$ as the advanced time. Furthermore, it is worthwhile to note that the quantity $\Phi$, usually called as the energy density of the radiation flux, does not have a direct operational meaning because the tangent null vector $k_{\alpha}$ does not have a natural normalization. Thus, it is preferable, see also \cite{Lindquist1965}, to consider the following quantity:
\begin{equation}
    \rho = \Phi (k_{\alpha}u^{\alpha})^{2},
\end{equation}
which defines the energy density as measured locally by an observer with a timelike 4-velocity $u^{\alpha}$. 

\subsection{Incompleteness of the EFL Coordinates} \label{proof_incompleteness}

\begin{figure}[htp]
\includegraphics[width=0.5\textwidth]{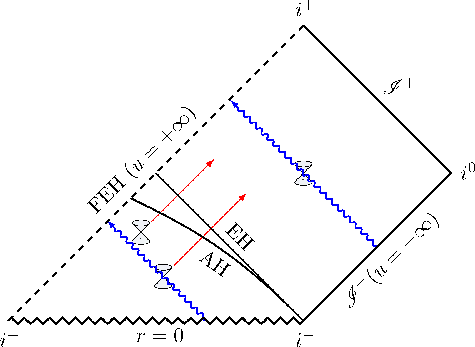}
\caption{The Penrose diagram \cite{penrose2011republication} of the outgoing $(\epsilon = -1)$ Vaidya metric, the coordinate $u$ in the EFL coordinates is indicated on the diagram, with the electromagnetic radiation is shown by red straight lines and the paths of incoming radial null geodesic (single photons) are shown by blue jagged lines. Notably, the apparent horizon (AH) and the event horizon (EH) do not coincide in the future, but they must have been the same hypersurface in the past.}
\label{fig:Incomp_EFL}
\end{figure}

In this subsection, we demonstrate why the EFL coordinates $(u,r,\theta,\phi)$ do not provide a complete description of the Vaidya manifold. The incompleteness of these coordinates is the primary motivation for the search for new coordinates in which the manifold is complete, allowing radial null geodesics to continue moving to infinite values of their affine parameter or be terminated upon encountering a gravitational singularity. The incompleteness of the coordinates $(u,r,\theta,\phi)$ becomes evident when studying the behavior of the ingoing radial null geodesics, emanating from the past null infinity $\mathscr{I}^{-}$ or from the past singularity surface $r=0$, for the case $\big(0<m(\infty)<\infty \big)$. It was suggested, but not proven in \cite{Israel1967, Fayos_1995}, that the geodesics appear to approach the future event horizon (FEH) surface, $r=2m(\infty)$, as $u \rightarrow \infty$, though they actually reach it for finite values of their affine parameter, see Fig. \ref{fig:Incomp_EFL}.

To support these insightful claims, we present a more articulated proof. We draw attention to the fact that, whereas Fig. \ref{fig:Incomp_EFL} is only valid for outgoing radiation, the forthcoming proof is valid for both ingoing and outgoing radiation. Let us consider the two branches of radial null curves, for which $ds^2=0$ and $\theta = \phi = \text{const}$. The first branch is given by $u=\text{const}$ (red), and the second branch (blue) is given by the solution of the following ordinary differential equation \footnote{This differential equation is classified as Abel second type class A \cite{Kamke}, which does not have a general solution.},
\begin{equation}
\label{second_branch}
\frac{du}{dr} =\frac{2 \epsilon r}{r-2m(u)}.
\end{equation}
We assume the following to hold
\begin{equation}
  0 < m(\pm \infty)< \infty,
\end{equation}
the question now arises as to whether the affine parameter $\lambda$ remains finite as $r \rightarrow 2m(\pm \infty)$ along the second branch. In order to answer this question we write the second branch (\ref{second_branch}) as a system of $1^\text{st}$ order ODEs
\begin{align}
  \label{dot_r}
  \dot{r} &= \frac{r-2m(u)}{\lambda},\\
  \dot{u} &=  \frac{2\epsilon r}{\lambda},
  \label{dot_u}
\end{align}
where an overdot indicates $d/d\lambda$, so that differentiation of the previous system with respect to $\lambda$ produces the geodesic equations of (\ref{Vaidya_EF_coords})
\begin{align}
  \label{2_dot_r}
  \ddot{r} &= - \frac{4 \epsilon m^{'}(u)r}{\lambda^2},\\
  \ddot{u} &= - \frac{4\epsilon m(u)}{\lambda^2},
  \label{2_dot_u}
\end{align}
where use has been made of both (\ref{dot_r}) and (\ref{dot_u}). Now let us assume that $\lambda \rightarrow \pm \infty$ as $r \rightarrow 2m(\pm \infty)$ then by virtue of (\ref{dot_u}) and (\ref{2_dot_u}) we obtain
\begin{equation}
\lim_{\lambda\rightarrow \pm \infty} \dot{u}=  \lim_{\lambda\rightarrow \pm \infty} \ddot{u} = 0,
\end{equation}
which is not possible as this changes the second geodesic branch into the first \footnote{Note that the first branch is characterized by $u=\text{const}$, which entails $\dot{u} = \ddot{u} = 0$.}. Therefore, our assumption is wrong, and we conclude that $\lambda$ along the second branch remains finite as $r \rightarrow 2m(\pm\infty)$. If we write this value of $\lambda$ as $\lambda_{0}$, we obtain
\begin{equation}
\label{radial_v}
  \lim_{\lambda\rightarrow \lambda_{0}} \dot{r} = 0,
\end{equation}
and
\begin{equation}
  \lim_{\lambda\rightarrow \lambda_{0}} \dot{u} =  \frac{4\epsilon m(\pm \infty)}{\lambda_0}.
\end{equation}
Evidently, the last equation remains finite because the mass function $m(\pm \infty)$ is assumed finite from the beginning. By virtue of  (\ref{radial_v}), we conclude that the region $\big(r<2m(\pm \infty)\big)$ is inaccessible in the EFL coordinates. Therefore, an extension is necessary. 
\section{Israel Coordinates} \label{Sec:Israel}
 In order to overcome the ``incompleteness problem" of the EFL coordinates, Israel \cite{Israel1967} introduced what he described as the analytic completion of the Vaidya manifold (\ref{Vaidya_EF_coords}). In Israel coordinates $(u,w,\theta,\phi)$, the Vaidya line element reads 
\begin{equation}
\label{Israel_original}
\textstyle
  ds^{2} = \big(\frac{w^2}{2m(u)r(u,w)}+\frac{4m^{'}(u)}{U(u)}\big) du^2+2dudw+r(u,w)^{2}d\Omega^{2}_{2},
\end{equation}
where $U(u) = \int_{0}^{u} \frac{du}{4m(u)}$, $r(u,w) = U(u)w+2m(u)$, and the function $m(u)$ is always positive. Notice that (\ref{Israel_original}) suffers a true singularity at $r(u,w) = 0$, see (\ref{1Weyl_Inv}), and at $u=0$, if $m'(u)$ does not vanish there, as explained below. To avoid any possible confusion about what is to be said, let us label the EFL retarded coordinate, $u$, as $t$. This then shows that (\ref{Israel_original}) is reduced to the outgoing Vaidya metric, (\ref{Vaidya_EF_coords}) with $u=t$ and $\epsilon=-1$, by the transformation 
\begin{equation}
    t(u) = -\int_{0}^{u} \frac{du}{U(u)},
\end{equation}
regular for ($u>0$, $t<\infty$). Apart from the cumbersome nature of Israel coordinates, the Vaidya metric in Israel coordinates (\ref{Israel_original}) does not adequately represent both the internal and external fields as long as the mass function $m(u)$ is only defined for $u \geq 0$. Since $u=0$ corresponds to $t=+\infty$ $\big(t(u)\propto -\log U(u)\big)$, it is impossible to extend the line element to the range $(u<0)$ via a coordinate transformation, as it would require knowledge of the mass function $m(t>\infty)$, i.e., beyond FEH. Hence, we believe that the ``maximal" extension of the Vaidya manifold, as given by the line element (\ref{Israel_original}), is imprecise. It is worth noting that there was an attempt \cite{Fayos_1995} to extend the Vaidya metric in terms of Israel coordinates. However, this approach faced the same problem as the original Israel extension of relying on coordinate transformations and the necessity of knowing the mass function $m(u)$ beyond the FEH in advance. It is also worthy of notice that although Israel coordinates have obvious advantages over the EFL coordinates, the Vaidya metric in Israel coordinates has not gained enough attention. To our knowledge, the metric has only been used once (see \cite{PhysRevD.31.233}) to study the complete gravitational collapse of a radiating shell of matter. Prior to the attempt given in \cite{PhysRevD.31.233}, all the work done to investigate the gravitational collapse in the presence of radiation was not complete. That is, the gravitational collapse was not followed beyond the event horizon because the Vaidya manifold in the EFL coordinates only describes the external field around a collapsing radiating object.   
\section{General Coordinate Construction} \label{Sec:Construction}
Consider the following general spherically symmetric metric expressed in the coordinates $(u,w,\theta,\phi)$ \cite{Lake_2006} 
\begin{equation}
  \label{general_metric}
ds^{2} = f(u,w) du^{2}+2h(u,w) du dw + r(u,w)^{2}d\Omega^{2}_{2},
\end{equation}
where $r(u,w)$ measures the area of the $2$-sphere $u=w=\text{const}$. The energy momentum tensor is once more taken to be of the eikonal form,
\begin{equation}
  \label{Vaidya_EMT}
T^{\alpha\beta} = \Phi k^{\alpha}k^{\beta},
\end{equation}
where $k^{\alpha} = \delta^{\alpha}_{w}$ is a radial null vector and the quantity $\Phi(k^{\alpha}u_{\alpha})^2$ is the energy flux, measured by an observer with tangent $u_{\alpha}$. Straightforward calculations \cite{grtensoriii} show that the only non-zero component of the Einstein tensor is $G^{w w}$ from which $\Phi$ can be directly obtained. If we take radial null trajectories with four-tangent $k^{\alpha}$ to be radial null geodesics affinely parametrized by $w$, i.e.,
\begin{equation}
    k^{\beta} \nabla_{\beta}k^{\alpha} = 0,
\end{equation}
this yields 
\begin{equation}
    \partial h(u,w)/\partial w = 0. 
\end{equation}
Thus, the function $h(u,w)$ reduces to a function of only $u$, $h(u,w)\equiv h(u)$. While we will limit ourselves to the choice $h(u) = \pm1$, we will keep the function as is for potential future use.
\subsection{Solving the Einstein Field Equations}
First \footnote{This approach of solving the field equations was first introduced in \cite{Lake_2006} to express the Schwarzschild-de Sitter vacuum metric in Israel coordinates, and was later utilized in \cite{Keenan} to obtain the Vaidya metric in the same set of coordinates.}, we benefit from the vanishing of the $G^{uu}$ component to obtain
\begin{equation}
\frac{\partial ^{2}}{\partial w^{2}} r(u,w)= 0.
\end{equation}
This leads, by integration, for a general expression \footnote{We also note that this expression can be deduced by assuming that (\ref{general_metric}) has a vanishing second Ricci invariant \cite{bisson2023israel}. This result is of interest because it is obtained directly from the geometry of the spacetime before considering the matter content. However, in order to show that (\ref{r_expression}) is the unique solution, further information must be given. Here the information is given by $G^{uu} = 0$. In \cite{bisson2023israel} the information is provided by the existence of Killing vectors.
}, to $r(u,w)$
\begin{equation}  \label{r_expression}
r(u,w) = f_1(u)w+f_2(u).
\end{equation}
In the sequel all the functions $f_n (u)$ are assumed suitably smooth \footnote{ All the functions are assumed to be at least $C^{2}$.}. Second, by solving $G^{\theta \theta} = 0$, with the aid of (\ref{r_expression}), we obtain
\begin{equation}
  \label{f_pde_eq}
  \textstyle
r(u,w)\frac{\partial ^{2}}{\partial w^{2}} f(u,w) + 2f_1(u)\frac{\partial }{\partial w}f(u,w) - 4h(u)\frac{d }{du}f_1(u) =0.
\end{equation}
Integrating (\ref{f_pde_eq}) yields
\begin{equation} \label{f_expression}
    \begin{split}
    f(u,w)= &\frac{2 f_1^{'}(u) h(u) f_2(u)^{2}-f_1(u)f_3(u)}{f_1(u)^{2}r(u,w)}\\
    &+\frac{2 f_1^{'}(u) h(u)w}{f_1(u)}+f_4(u),
\end{split}
\end{equation}
where $(')$ denotes ordinary differentiation with respect to the coordinate $u$. By solving $ G^{uw} = 0 $, we find that $f_4(u)$ is given by
\begin{equation}
  \label{f4_expression}
f_4(u) = \frac{h(u)\big(2f_{1}(u)f_2^{'}(u)-h(u)\big)}{f_1(u)^2},
\end{equation}
where use has been made of (\ref{r_expression}) and (\ref{f_expression}). By virtue of (\ref{r_expression}), (\ref{f_expression}), and (\ref{f4_expression}) the only non-zero component of the Einstein tensor can be given as 
\begin{equation}\label{G^ww_component}
  \begin{split}
   G^{ww} = &\frac{1}{\chi(u)}\bigg(2h(u)^2f_2(u)^2f_1^{''}(u)+4h(u)^2f_2(u)f_1^{'}(u)f_2^{'}(u) \\ &\qquad-h(u)f_3(u)f_1^{'}(u)-2h(u)f_2(u)^{2} h^{'}(u)f_1^{'}(u)\\ &\qquad -h(u)  f_1(u)f_3^{'}(u)+2f_1(u)f_3(u)h^{'}(u) \bigg),
\end{split}
\end{equation}
where $\chi(u,w)=h(u)^4f_1(u)r(u,w)^2$. The $G^{ww}$ is conveniently expressed in the following way. First define the Hernandez-Misner mass \cite{Hernandez1966}
\begin{equation} \label{mass_def}
m \equiv \frac{r(u,w)^3}{2} R_{\theta \phi}^{\quad\theta \phi},
\end{equation}
where $R$ is the Riemann tensor. By calculating $R_{\theta \phi}^{\quad\theta \phi}$ for (\ref{general_metric}) and making the necessary simplifications, (\ref{mass_def}) can be given in terms of the characterizing functions $f_n(u)$ as
\begin{equation}
  \label{H_M_mass}
m = m(u) =  \frac{2h(u)f_2(u)^2f_1^{'}(u)-f_1(u)f_3(u)}{2h(u)^{2}},
\end{equation}
where the mass function must always remain positive-valued over its domain. As a result, $G^{ww}$ can be expressed in a more succinct form,
\begin{equation}
  \label{Phi_expression}
G^{ww} = \frac{2 m^{'}(u)}{h(u)f_1(u)r(u,w)^2} = 8 \pi \Phi.
\end{equation}
Similarly, a more convenient expression of the function $f(u,w)$ can be obtained with the aid of (\ref{r_expression}), (\ref{f_expression}), (\ref{f4_expression}), and (\ref{H_M_mass})
\begin{equation}
\label{f_expression_1}
f(u,w) = \frac{\mathcal{A}(u) r(u,w)^2 +\mathcal{B}(u) r(u,w)+\mathcal{C}(u)}{f_1(u)^2r(u,w)},
\end{equation}
where
\begin{align} 
\mathcal{A}(u) &= 2h(u)f_1^{'}(u),\\
\mathcal{B}(u) &= 2h(u)f_1(u)f_2^{'}(u)-2h(u)f_2(u)f_1^{'}(u)-h(u)^2,\\
\mathcal{C}(u) &= 2h(u)^2m(u).
\end{align}
\section{Physical Restrictions on the Choice of the Characterizing Functions} \label{Sec:Restrictions}
The first restriction that we impose, using (\ref{H_M_mass}), is given by the following inequality 
\begin{equation}\label{prop1}
2h(u)f_2(u)^2f_1^{'}(u)>f_1(u)f_3(u).
\end{equation}
This is necessary to ensure that the mass function, $m(u)$, is always positive.
The second restriction is that the measured radiation flux is a positive quantity,
\begin{equation} \label{WEC_eq}
    \Phi (k^{\alpha}u_{\alpha})^2> 0.
\end{equation}
Substituting (\ref{Phi_expression}) in (\ref{WEC_eq}) and simplifying, we obtain 
\begin{equation} \label{WEC_ineq}
    \frac{m^{'}(u)}{h(u)f_1(u)}>0,
\end{equation}
which dictates that the signs of $m^{'}(u)$ and $h(u)f_{1}(u)$ have to be identical. As our attention is confined to classical matter fields (radiation), a minimum requirement is that this matter distribution must satisfy the Weak Energy Condition (WEC). This requirement implies, with the aid of (\ref{Phi_expression}), the following stipulations on the different forms of radiation, summarized in Table \ref{table.1}.

 \begin{table}[hbt!]
  \caption{\label{table.1}
Stipulations on the functions $h(u)$ and $f_1(u)$.}
 \begin{ruledtabular} 
\begin{tabular}{l c c c}
                 Direction                          & $m'(u)$                 & $h(u)$     & $f_1(u)$   \\ \hline 
\multirow{2}{*}{Outgoing Radiation} & \multirow{2}{*}{$m'(u)<0$}\qquad & $h(u)<0$   & $f_1(u)>0$ \\ 
                                             &                         & $h(u) > 0$ & $f_1(u)<0$ \\ 
\multirow{2}{*}{Ingoing Radiation}  & \multirow{2}{*}{$m'(u)>0$}\qquad & $h(u)>0$   & $f_1(u)>0$ \\ 
                                             &                         & $h(u)<0$   & $f_1(u)<0$ 
\end{tabular}
\end{ruledtabular}
\end{table}
Table. \ref{table.1} clearly illustrates that both ingoing and outgoing radiation can be obtained without changing the sign of the function $h(u)$. However, as will be seen shortly, the direction of radiation in the EFL coordinates is dictated by the sign of the function $h(u)$.
\section{The Apparent Horizon and the Event Horizon} \label{Sec:Horizons}
We begin this section by providing a general derivation to the location of the apparent horizon of (\ref{general_metric}). To this end, let us examine the congruence of radial null trajectories 
 characterized by the four-tangent $\ell^{\alpha}$,
\begin{equation}
\ell^{\alpha} = \delta^{\alpha}_{u}-\frac{f(u,w)}{2h(u)}\delta^{\alpha}_{w},
\end{equation}
However, it does not satisfy the geodesic equation in the affine-parameter form. This is evident from the equations $\ell^{\alpha}\nabla_{\alpha}\ell^{u} = \kappa \ell^{u}$ and $\ell^{\alpha}\nabla_{\alpha}\ell^{w} = \kappa \ell^{w}$, where $\kappa = \kappa (u,w)$ and it is called the inaffinity. The geodesics equations are:
\begin{equation}
 \ell^{\alpha}\nabla_{\alpha}\ell^{u} = \left(\frac{2\frac{d}{d u}h(u)-\frac{\partial}{\partial w}f(u,w)}{2h(u)}\right)(1) = \kappa \ell^{u},    
\end{equation}
and
\begin{equation}
  \ell^{\alpha}\nabla_{\alpha}\ell^{w} = \left(\frac{2\frac{d}{d u}h(u)-\frac{\partial}{\partial w}f(u,w)}{2h(u)}\right)\left(-\frac{f(u,w)}{2h(u)} \right) = \kappa \ell^{w},  
\end{equation}
with the inaffinity $\kappa$ given by
\begin{equation}\label{kappa_1}
    \kappa = \frac{2\frac{d}{du}h(u)-\frac{\partial}{\partial w}f(u,w)}{2h(u)}.
\end{equation}
The associated expansion scalar $\Theta^{(\ell)}$ of this non affinley parametrized congruence of radial null geodesics, see \cite{Poisson2004,Blau_notes} for the definition of the expansion in this case, is given by 
 \begin{align} \label{expansion_l}
\Theta^{(\ell)} &= \nabla_{\alpha}\ell^{\alpha}-\kappa,\nonumber \\
       &= -\frac{r\left(u,w\right) \frac{\partial}{\partial w}f \left(u,w\right)-2 r\! \left(u,w\right) \frac{d}{d u}h (u)}{2 h \left(u\right) r\! \left(u,w\right)} \nonumber \\
       &\quad - \frac{2 f (u,w) \frac{\partial}{\partial w}r \left(u,w\right)-4 h \left(u\right) \frac{\partial}{\partial u}r \left(u,w\right)}{2 h \left(u\right) r\! \left(u,w\right)}-\kappa,\nonumber \\
       &= \textstyle - \frac{1}{h(u)r(u,w)} \big( f(u,w) \frac{\partial}{\partial w}r(u,w)-2h(u)\frac{\partial }{\partial u}r(u,w)\big).
\end{align}
The apparent horizon is characterized by $\Theta^{(\ell)} = 0$, and thus by virtue of (\ref{expansion_l}) we obtain the following condition
\begin{equation}
\label{AH_condition}
    2h(u)\frac{\partial r(u,w)}{\partial u} = f(u,w) \frac{\partial r(u,w)}{\partial w}.
\end{equation}
We substitute (\ref{r_expression}) in (\ref{AH_condition}), which yields
\begin{equation}
    2h(u) \left( f_1^{'}(u)w+f_2^{'}(u)\right) = f(u,w)f_1(u).
\end{equation}
With the aid of (\ref{f_expression_1}) the previous equation takes the form 
\begin{equation}
\begin{split}
0&= 2f_1^{'}(u)r(u,w)^2+2h(u)m(u)\\ 
& \quad -\left( 2w f_1(u)f_1^{'}(u)+2f_2(u)f_1^{'}(u)+h(u)\right)r(u,w).
\end{split}
\end{equation}
We can use (\ref{r_expression}) once more to reduce the last equation to 
\begin{equation}
    -h(u)\big( r(u,w)-2m(u) \big) = 0,
\end{equation}
which immediately gives the sought-after result:
\begin{equation} \label{General_result_AH}
    r(u,w) = 2m(u).
\end{equation}
It is thus established that the apparent horizon is located at $r=2m(u)$.
We also note that the previous result is established before making any choices for the characterizing functions, $f_n(u)$. Determining the location of the event horizon in the Vaidya metric is not as straightforward as locating the apparent horizon. In fact, the entire future history of the metric, as specified by the functions $f(u,w)$ and $h(u)$, must be predetermined in order to identify the null generators of the event horizon \cite{Poisson2004}.
However, we may generically define the future (past) event horizon as a causal boundary for the timelike geodesics terminating at future (past) timelike infinity, $i^{+}(i^{-})$ \footnote{For the definitions of these infinities we refer to \cite{penrose2011republication}.}.
\section{Specific Coordinate Representations of the Vaidya Metric}\label{Sec:Representations}
In this section, we demonstrate that we can obtain various coordinate representations of the Vaidya metric by selecting different expressions for the characterizing functions, $h(u)$ and $f_n(u)$. Additionally, we emphasize that the meaning of the coordinate $u$ is dependent on the choice of the characterizing functions, and thus the coordinate $u$ in the EFL coordinates has a different interpretation to that in Israel coordinates. 
\subsection{The Vaidya Metric in the EFL Coordinates}
Let us choose the characterizing functions such that $h(u) = \pm 1$, $f_1(u) = 1$, and $f_2(u) = 0$, then we obtain $w = r$ with the help of (\ref{r_expression}). Furthermore, we get $f_3(u) = -2m(u)$ from  (\ref{H_M_mass}). Substituting these values in (\ref{f_expression_1}) yields
\begin{equation}
f(u,r) = \frac{-r+2m(u)}{r},
\end{equation}
and thus the metric (\ref{general_metric}) becomes
\begin{equation}
  \label{Vaidya_coordinates}
ds^2 = -\Big(1-\frac{2m(u)}{r}\Big)du^2\pm 2dudr+r^2d\Omega_{2}^2,
\end{equation}
with $G^{ww} = \pm \frac{ 2m^{'}(u)}{r^2}$. It is clear that, with the help of Table \ref{table.1}, we can obtain $h(u) = -1$ for the outgoing radiation version of the Vaidya metric, where the coordinate $u$ is a retarded time. Similarly, selecting $h(u) = +1$ yields the ingoing radiation version of the Vaidya metric, with $u$ as an advanced time.
\subsection{The Vaidya Metric in Israel Coordinates} \label{Israel_section}
In this subsection, we explore how by introducing different choices to the functions $f_n(u)$, we obtain Israel coordinates. Let us consider the following choices: $f_1(u) = U(u)$, $f_2(u) = 2 M(u)$, and $f_3(u) = 0$. It follows from (\ref{H_M_mass}) that for $M(u)=m(u)$ (which is a choice),
\begin{equation}
    U^{'}(u) = \frac{h(u)}{4m(u)}.
\end{equation}
 Thus, with the aid of the first fundamental theorem of calculus we write
\begin{equation}\label{U_def}
U(u) = \int_{0}^{u} \frac{h(x)}{4m(x)} dx.
\end{equation}
However, since our choices for the function $h(u)$ will be confined to either $+1$ or $-1$, we set $h(u)=h=\pm1$. Consequently, the expression (\ref{U_def}) takes the form 
\begin{equation}
    \label{U_def_simplified}
U(u) = h\int_{0}^{u} \frac{1}{4m(x)} dx.
\end{equation}
It follows that the spacetime line element (\ref{general_metric}) can be written as
\begin{equation}
  \label{Israel_ext_general}
  ds^{2} = \left(\frac{w^2}{2m(u)r}+\frac{4hm^{'}(u)}{U(u)}\right) du^2+2hdudw+r^{2}d\Omega^{2}_{2},
\end{equation}
where $r$  is no longer a coordinate; it is now a function $r=r(u,w) = U(u)w+2m(u)$ and $G^{ww} = \frac{2hm^{'}(u)}{U(u)r(u,w)^2}$. Here, $u$ is a null coordinate and (\ref{Israel_ext_general}) describes both outgoing and ingoing radiation. It is interesting to note that  the presence of $h$ is not necessary for (\ref{Israel_ext_general}), as demonstrated in \cite{Lake_2006}, particularly when $m^{'}(u)=0$. It is noteworthy that, in accordance with (\ref{General_result_AH}), the apparent horizon is now located at $w=0$. 
There is some ambiguity regarding the sign of $u$ which appears in the definition of the function $U(u)$ (\ref{U_def_simplified}); for example, in \cite{Israel1967}, $u$ is always positive, whereas in \cite{Fayos_1995} $u$ can be either positive or negative. We shall resolve this ambiguity and demonstrate when $u$ can be negative or positive. To this end, recall that  
\begin{equation}
    U^{'}(u) = \frac{h}{4m(u)},
\end{equation}
which means that the sign of  $U^{'}(u)$ is solely determined by the sign of $h$. Also, with the aid of the WEC, (\ref{WEC_ineq}), and (\ref{U_def_simplified}), we have
\begin{equation}
    \frac{m^{'}(u)}{hU(u)} = \frac{m^{'}(u)}{\int_{0}^{u}\frac{dx}{4m(x)}} > 0,
\end{equation}
where in the last equation we have taken $h^2 = 1$. Hence, for $m^{'}(u)>0$ the integral must be positive ($u$ in the integral must be positive) and for $m^{'}(u)<0$ the integral has to be negative ($u$ in the integral must be negative). Consequently, we have seen that the sign of $u$ in the integral is not always positive like in \cite{Israel1967}, and the dichotomy in the function $U(u)$ based on the sign of $u$ is explained in a more articulated way. We have summarized all the choices we have considered thus far in Table \ref{Israel_table}.
\begin{table*}
\caption{\label{Israel_table} A summary of the different choices of the characterizing functions that appear in our construction alongside with the resulting metrics.}
\setlength\extrarowheight{5pt}
\begin{ruledtabular}
\begin{tabular}{c c c c c}
$h$& $f_1(u)$   & $f_2(u)$ & $f_3(u)$ & The Resulting Metric  \\  \hline                                                                        
$-1$            & $+1$    & $0$               & $-2m(u)$          & Outgoing Vaidya (EFL)    \\                                                                                                            
$+1$            & $+1$    & $0$               & $-2m(u)$          & Ingoing Vaidya (EFL)    \\                                                                                                          
$+1$            & \quad$U(u)=\int_{0}^{u<0}\frac{dx}{4m(x)} = \int_{0}^{u>0}\frac{-dx}{4m(x)}$                                               & $2m(u)$           & $0$               & Outgoing Israel    \\ 

$-1$            & \quad$U(u)=\int_{0}^{u<0}\frac{-dx}{4m(x)} =  \int_{0}^{u>0}\frac{dx}{4m(x)} $
                                & $2m(u)$           & $0$               & Outgoing Israel   \\ 

$+1$            & \quad$U(u)=\int_{0}^{u>0}\frac{dx}{4m(x)}  = \int_{0}^{u<0}\frac{-dx}{4m(x)}$& $2m(u)$          & $0$               & Ingoing Israel    \\ 
$-1$            & \quad$U(u)=\int_{0}^{u>0}\frac{-dx}{4m(x)} =  \int_{0}^{u>0}\frac{-dx}{4m(x)}$   & $2m(u)$           & $0$               & Ingoing Israel  \\ 
\end{tabular}
\end{ruledtabular}
\end{table*}
Finally, we introduce a restriction on the $w$ coordinate corresponding to the the surface $r(u,w) = 0$, the physical singularity, see below. Since $r(u,w) = U(u)w+2m(u)$, for $r(u,w) = 0$ we obtain
\begin{equation}
  w = -\frac{2m(u)}{U(u)} \equiv w_{0}(u),
\end{equation}
and so $w_{0} > 0$ for $U(u)<0$ and $w_{0} < 0$ for $U(u)>0$. It turns out that this exactly the case when we study the radial null geodesics in the proposed maximal extensions of the Vaidya metric \cite{Null_Israel_Completeness}. 
\section{Invariants} \label{Sec:Invariants}
 Up to syzygies \cite{SANTOSUOSSO1998381}, we find that the only non-differential non-vanishing invariant of (\ref{general_metric}) is the first Weyl invariant,
 \begin{align} \label{Weyl_Inv_gen}
w1R &\equiv  \frac{1}{8}C_{\alpha \beta \gamma \delta}C^{\alpha \beta \gamma \delta}\nonumber\\  
            &= \frac{3}{2h(u)^4r(u,w)^6}\big(f_1(u)f_3(u)-2h(u)f_1(u)'f_2(u)^2\big),
\end{align}
 which reduces to the following expression in Israel coordinates,
\begin{equation} \label{1Weyl_Inv}
w1R \equiv  \frac{1}{8}C_{\alpha \beta \gamma \delta}C^{\alpha \beta \gamma \delta}  = \frac{6m(u)^2}{r(u,w)^6},
\end{equation}
where $C_{\alpha \beta \gamma \delta}$ is the Weyl tensor. However, as (\ref{Phi_expression}) makes clear, it would be informative to have invariant information for $m^{'}(u)$. This is obtained by way of the Bach tensor \cite{Bach_tensor}, see also \cite{Glass}. First define
\begin{equation}
  A_{\alpha\beta\delta} = \nabla^{\gamma}C_{\alpha\gamma\beta\delta},
\end{equation}
where $\nabla^{\gamma}$ denotes contravariant derivative. The Bach tensor is given by 
\begin{equation}
  B_{\alpha \beta} = \nabla^{\delta} A_{\alpha \beta\delta}+\frac{R^{\gamma \delta}C_{\alpha\gamma\beta\delta}}{2}.
\end{equation}
Since the Bach tensor is trace-free, the first Bach invariant is
\begin{equation}
  B\equiv B_{\alpha\beta}B^{\alpha\beta}.
\end{equation}
In the present case we find, with the aid of (\ref{Phi_expression}), that 
\begin{equation}
  B = \left(\frac{4U(u)m^{'}(u)}{r(u,w)^4}\right)^2.
\end{equation}
Nevertheless, the preceding result does not provide the desired invariant definition of $m'(u)$ due to its dependence on the functions $r(u,w)$ and $U(u)$. 
\section{Summary and Discussion} \label{Sec:Summary}
We have examined the construction of Israel coordinates for the Vaidya metric and have simplified the problem to finding appropriate expressions for the characterizing functions that arise from integrating the field equations. This construction is systematic and does not necessitate any coordinate transformation, which provides us with the chance to spot potential extensions of the Vaidya manifold by introducing distinct expressions for the characterizing functions, $f_n(u)$. Nonetheless, the main focus of this paper is to reconstruct Israel coordinates for the Vaidya metric. By utilizing the WEC, we have understood the role of the function $h(u)$ in the Vaidya metric. Although the sign of the $h(u)$ is paramount in determining the direction of radiation in the EFL coordinates, we have demonstrated that this is not the case for Israel coordinates. That is, both ingoing and outgoing radiation can be achieved with $h=+1$ or $h=-1$. However, the impact of changing the sign of the function $h(u)$ will be further investigated  when we discuss the completeness of Israel coordinates in \cite{Null_Israel_Completeness}. 
\section{Acknowledgement}
This work was supported (in part) by a grant from the Natural Sciences and Engineering Research Council of Canada (to KL). 
\bibliography{PaperI_bib}
\end{document}